\documentclass[preprint,aps,floatfix]{revtex4-1}
\usepackage{hyperref}
\everymath={\displaystyle}
\usepackage{young}
\usepackage{rotating}
\usepackage{longtable}
\def\1{\mbox{l\hspace{-0.53em}1}}
\newcommand{\fr}{\frac}
\usepackage{multirow}

\newlength{\AccoHaut}

\begin{document}
\title{Updated $1/N_c$ expansion analysis of $[{\bf 56, 2^+}]$ and  $[{\bf 70, \ell^+}]$ baryon multiplets}

\author{N. Matagne\footnote{E-mail address: nicolas.matagne@umons.ac.be}}
\affiliation{University of Mons, Service de Physique Nucl\'eaire et 
Subnucl\'eaire,Place du Parc 20, B-7000 Mons, Belgium}

\author{Fl. Stancu\footnote{E-mail address: fstancu@ulg.ac.be}}
\affiliation{University of Li\`ege, Institute of Physics B5, Sart Tilman,
B-4000 Li\`ege 1, Belgium}

\date{\today}

\begin{abstract}

The mass spectra of the
$[{\bf 56, 2^+}]$ and $[{\bf 70, \ell+}]$ multiplets, both belonging to the $N$ = 2 band,
is reviewed in the $1/N_c$ expansion method. 
Previous studies, separately made for each multiplet, are presently updated to the 2014 Particle Data Group.
The mass formula including corrections up to $\mathcal{O}(1/N_c)$ and first order in SU(3) flavor symmetry breaking, 
has the same independent operator basis in both cases.
A special emphasis is made on the role of the
SU(3) symmetry breaking operators $B_i$ $(i = 1,2,3)$.
This can allow for multiplet assignment of $\Lambda$ and $\Sigma$ hyperons, which generally is quite difficult to make.
Tentative assignments of hyperons with two- and one-star resonances are made to the  $[{\bf 70, \ell+}]$ multiplet.
Another important aim is to find out whether or not a common value of the coefficient $c_1$ of the dominant operator
in the mass formula, can well fit 
the present data in both multiplets. A negative answer, which is here the case, implies distinct 
Regge trajectories for symmetric and mixed symmetric states.

\end{abstract}

\maketitle

\section{Introduction}

The $1/N_c$ expansion method proposed by 't Hooft \cite{'tHooft:1973jz} and applied to baryons 
by Witten  \cite{Witten:1979kh}, where $N_c$ is the number of colors,
is based on the discovery
that, for $N_f$ flavors, the ground state baryons
display an exact contracted SU($2N_f$) spin-flavor symmetry in the
large $N_c$ limit of QCD \cite{Gervais:1983wq,DM93}. 
The Skyrme model, the strong coupling theory \cite{Cook:1965qu} and the static quark model share a common
underlying symmetry with QCD baryons in the large $N_c$ limit \cite{Bardakci:1983ev}.

Presently the $1/N_c$ expansion method is considered to be a model independent, powerful and  systematic tool for
baryon spectroscopy. 
It has been applied with great success
to the ground state baryons ($N = 0$ band), described
by the symmetric representation $\bf 56$ of SU(6) 
\cite{DM93,Jenk1,DJM94,DJM95,CGO94,JL95,DDJM96}. 
At $N_c \rightarrow \infty$ the ground state baryons are degenerate.
At large, but finite $N_c$, the mass splitting 
starts at order $1/N_c$ as first observed in Ref. \cite{Bardakci:1983ev}.

The extension of the $1/N_c$ expansion method to excited states 
requires the symmetry group  SU($2N_f$) $\times$ O(3) \cite{Goi97}, in order to introduce orbital excitations.
There is no $\textit {a priori}$  justification for this symmetry.
However, in practice, the experimentally observed resonances can approximately be classified  
as  SU($2N_f$) $\times$ O(3) multiplets, grouped into 
excitation bands, $N $ = 1, 2, 3, ..., 
each band containing a number of SU(6) $\times$ O(3) multiplets,
as  is done in quark models. In addition, lattice QCD studies 
have shown that the number of each spin and flavor states in the lowest energy bands is in agreement
with the expectations based on a weakly broken SU(6) $\times$ O(3) symmetry \cite{Edwards:2012fx}, used in quark models and in the 
treatment of excited states in large $N_c$ QCD,  as is here the case.

The extension of the $1/N_c$  expansion to excited states 
has been theoretically supported by Pirjol and Yan \cite{PY} who derived consistency conditions 
for excited states, similar to those for the ground state  \cite{Gervais:1983wq,DM93}.  Later on, the 
extension of the method to excited states was legitimated by Cohen and Lebed \cite{COLEB1}
who proved the compatibility between the meson-nucleon scattering picture
and the quark model type picture,
both leading to identical degeneracy patterns, giving rise to towers of states.

Some symmetric multiplets of SU(6) $\times$ O(3) classification, in particular $[{\bf 56}, 2^+]$ and $[{\bf 56}, 4^+]$,
containing two and four units of orbital excitations,
were analyzed by analogy to the ground state in Refs. \cite{GSS03} and \cite{Matagne:2004pm} respectively.
In this case the splitting starts at order $1/N_c$ as well.

The situation is technically more complicated for mixed symmetric states. Two approaches 
have been proposed so far. The first one is based on the Hartree approximation and
describes the $N_c$ quark system as a ground state symmetric core of $N_c - 1 $ 
quarks and an excited quark \cite{CCGL}. This implies the split of 
SU($2N_f$) generators into two parts, one acting on the  core and the other on the excited quark.
Naturally, the number of generators entering the mass formula becomes larger, hence the applicability
of the method beyond the $N$ = 1 band becomes more problematic  \cite{Matagne:2006zf}.

The second procedure, where the Pauli principle is
implemented to all $N_c$ identical quarks has been proposed in Refs. \cite{Matagne:2006dj,Matagne:2008kb}.
There is no physical reason to separate the excited quark from the rest of the system.
The method can straightforwardly be applied to all excitation bands $N$. It requires 
the knowledge of the matrix elements of all the  SU($2N_f$) generators acting on mixed 
symmetric states described by the partition $(N_c - 1,1) $.
In both cases the mass splitting starts at order $N^0_c$. A discussion on the comparison between 
the two methods and various applications can be found in Ref. \cite{Matagne:2014lla}. In the following we apply 
the procedure of Refs. \cite{Matagne:2006dj,Matagne:2008kb} to analyze baryons thought to belong to the
mixed symmetric $[{\bf 70, \ell^+}]$ multiplet and reanalyze the symmetric multiplet  $[{\bf 56}, 2^+]$,
which means that the same basis operators is used in the mass formula in both cases, which is a novel aspect
of this study.
     
In the next section we recall results previously obtained for multiplets in the $N$ = 2 band.
In Sec. \ref{massformula} we introduce the basis operators used in the mass formula and in Sec. \ref{matrixelements}
we derive or recall the analytic forms of the matrix elements of the basis operators.
Sec.  \ref{fit} is devoted to a numerical fit which gives the values of the dynamical coefficients
entering the mass formula followed by a discussion of the obtained resonance masses.  Conclusions are drawn
in the last section. In Appendix \ref{operatorO2} we recall the analytic formula of the single particle spin-orbit 
operator used for mixed symmetric $[{\bf 70, \ell^+}]$ multiplets.
In Appendix \ref{operatorO6} we present analytic details of the tensor operator $O_6$. In Appendix \ref{Mixing}
we discuss the mixing between spin quartets and doublets in a simplified model. 
In Appendix \ref{BREAK} we show details of the analytic calculations of the SU(3) flavor breaking
operators $B_2$ and $B_3$.

\section{Previous studies of the $N$ = 2 band in the $1/N_c$ expansion}

The $N$ = 2 band has the following multiplets
$[{\bf 56'},0^+]$,    $[{\bf 56},2^+]$, $[{\bf 70},0^+]$,  $[{\bf 70},2^+]$ and 
$[{\bf 20},1^+]$. The observed resonances are usually assigned to the symmetric [{\bf 56}] or the mixed symmetric $[{\bf 70}]$
SU(6) multiplets. The antisymmetric SU(6)  multiplet $[{\bf 20},1^+]$ has been ignored so far,
on the basis that it does not have a real counterpart. However, two new resonances $N(2100)1/2^+$* and $N(2040)3/2^+$*,
presently included in the 2014 Particle Data Group \cite{PDG}, 
were recently assigned to the $[{\bf 20},1^+]$ multiplet, by an educated guess, see Ref. \cite{Crede:2013kia} Table 23.
Before drawing any conclusion, stronger experimental evidence is required to confirm the existence of 
$N(2100)1/2^+$* and $N(2040)3/2^+$*.

The multiplet  $[{\bf 56'},0^+]$ describes states  with a radial excitation,
in particular the Roper resonance. It was the first to be studied in the large $N_c$  limit \cite{Carlson:2000zr},
by using a simplified mass formula of the  G\"ursey-Radicati type. The analysis was free of any assumption regarding the
wave function except its symmetry in SU(6). Strong decay widths were calculated as well.

The analysis of the  $[{\bf 56},2^+]$ baryon masses has first been  performed in Ref. \cite{GSS03}. It has been 
reconsidered in Ref. \cite{Matagne:2004pm} with nearly identical results and  the analysis has been extended to the higher multiplet   
$[{\bf 56},4^+]$ of the $N$ = 4 band in the same paper.

The $[{\bf 70},0^+]$ and $[{\bf 70},2^+]$  baryon masses were first analyzed in Ref. \cite{Matagne:2005gd} for $N_f$ = 2 and
extended in Ref. \cite{Matagne:2006zf} to $N_f$ = 3, both studies being performed within the symmetric core + excited
quark procedure \cite{CCGL}. 
The $[{\bf 70},\ell^+]$ ($\ell$ = 0, 2) multiplets were revisited \cite{Matagne:2013cca}
within the approach of Ref. \cite{Matagne:2006dj}
where the Pauli principle was fully taken into account.

In Refs. \cite{Matagne:2005gd} and \cite{Matagne:2013cca} Regge-type trajectories have been drawn for the most dominant
coefficient in the mass formula, denoted in the following by $c_1$ and somewhat conflicting results have been obtained.
The trajectories were drawn as a function of the band number $N = 0,1,2,3$ and 4.
While in Ref. \cite{Matagne:2005gd} a single trajectory has been obtained (note that large $N_c$ results for the $N$ = 3 band 
were not  available yet), in Ref. \cite{Matagne:2013cca} two distinct, nearly parallel, Regge trajectories have been obtained,
the lower one for symmetric $[{\bf 56}]$-plets and the higher one for mixed symmetric $[{\bf 70}]$-plets.

In this work we wish to clarify the issue, whether or not one or two distinct trajectories stem from the 
$1/N_c$ expansion, one for symmetric the other for mixed symmetric states.  
For this purpose we combine together the analysis of the $[{\bf 56}, 2^+]$ and $[{\bf 70}, \ell^+]$
multiplets of the $N$ = 2 band. An important aspect is that presently we use the same set of linearly independent
operators in the mass formula, which was not the case before. 
Details are given in the following sections.
We do not include the $[{\bf 56'},0^+]$ multiplet, associated with states having a radial excitation, 
because they can deteriorate the numerical fit, due to their 
location in the spectrum, too low from the other states.

Another incentive to perform this analysis was that the band number $N$ appeared to be a good quantum number 
for the spin-independent part of semirelativistic quark models \cite{Semay:2007cv,Semay:2007ff,Buisseret:2008tq}.
Therefore plotting $c_1^2$ as a function of $N$ seems meaningful, inasmuch as $c_1^2$ simulates the effect of the 
kinetic and the confinement parts of quark model Hamiltonians.

Presently we use the data of the 2014 Particle Data Group \cite{PDG}.
which sometimes give more precise values for the resonance masses with smaller error bars than before.
For example  $N(1720){\frac{3}{2}}^{+}$ has a mass of $1725\pm25$ MeV as compared to $1700\pm50$ MeV in 
the 2002 Particle Data Group \cite{PDG2002}, used in Ref. \cite{GSS03}.
The changes are due to a more complex analysis of all major photo-production of mesons in a coupled-channel   
partial wave analysis as done, for example, in Ref. \cite{Anisovich:2011fc}.

\section{The Mass Operator}\label{massformula}

The general form of the mass operator,  where the SU(3) symmetry is broken, has first been proposed in Ref. \cite{JL95} as
\begin{equation}
\label{massoperator}
M = \sum_{i}c_i O_i + \sum_{i}d_i B_i .
\end{equation} 
The operators $O_i$ are defined as the scalar products
\begin{equation}\label{OLFS}
O_i = \frac{1}{N^{n-1}_c} O^{(k)}_{\ell} \cdot O^{(k)}_{SF},
\end{equation}
where  $O^{(k)}_{\ell}$ is a $k$-rank tensor in SO(3) and  $O^{(k)}_{SF}$
a $k$-rank tensor in SU(2)-spin, but invariant in SU($N_f$).
Thus $O_i$ is rotational invariant.
For the ground state one has $k = 0$. The excited
states also require  $k = 1$  and $k = 2$ terms.
The $k = 1$ tensor has three components, which are the generators $L^i$ of SO(3). 
The components of the $k = 2$ tensor operator of SO(3) read \cite{Matagne:2005gd}
\begin{equation}\label{TENSOR} 
L^{(2)ij} = \frac{1}{2}\left\{L^i,L^j\right\}-\frac{1}{3}
\delta_{i,-j}\vec{L}\cdot\vec{L}.
\end{equation}

The operators $B_i$ break the SU(3) flavor symmetry and are defined to have zero expectation
values for nonstrange baryons. 

The angular momentum-independent operators up to $\mathcal{O}(N^{-1}_c)$ are 
\begin{equation}\label{Lindep}
O_1 = N_c \ \1 , ~~~  O_3 = \frac{1}{N_c} S \cdot S, ~~~  O_4 = \frac{1}{N_c} (T \cdot T - \frac{N_c(N_c+6)}{12}),
\end{equation}
where $S^i$, $T^a$ and $G^{ia}$ are the SU(6) spin, flavor and spin-flavor operators. The definition of $O_4$ 
where the term $N_c(N_c + 6)/12$ has been subtracted  is necessary for including SU(3) singlets  \cite{Matagne:2008kb}.
This definition gives the same matrix elements as the isospin operator $\frac{1}{N_c}T \cdot T$ restricted to nonstrange baryons 
within SU(4) symmetry \cite{Matagne:2006dj}. The form of $O_4$ is consistent with Eq. (5.12) of Ref. \cite{DJM95}.

The operators containing $O^{(k)}_{\ell}$ tensors are the spin-orbit operator $O_2$ and the tensor operator $O_6$, the latter 
being defined as
\begin{equation}\label{lop}
O_6 =  \frac{1}{N_c} L^{(2)ij} G^{ia} G^{ja}.
\end{equation}
For the $[{\bf 56}]$-plets the spin-orbit operator $O_2$  
is defined in terms of angular momentum $L^i$ components acting on the whole
system as in Ref. \cite{GSS03} and is order  $\mathcal{O}(1/N^c)$ 
\begin{equation}\label{LS}
 O_2 = \frac{1}{N_c} L \cdot S, ~~~   
\end{equation}
while for the $[{\bf 70}]$-plets it is defined as a single-particle operator $\ell \cdot s$ of order $\mathcal{O}(N^0_c)$, 
as used previously \cite{CCGL,Matagne:2006zf,Matagne:2013cca}.
Note that $O_6$ is normalized differently as compared to Ref. \cite{Matagne:2011fr}. Actually,
what matters in the mass formula is the product $c_i O_i$. Note also that $O_6$ is order  $\mathcal{O}(N^0_c)$ 
in the  $[{\bf 70}, 1^-]$ multiplet \cite{Matagne:2011fr} which is an important issue in the study of the
compatibility between the quark-shell picture and meson-nucleon scattering picture \cite{Matagne:2012vq}.
The existing compatibility legitimates the extension of the $1/N_c$ expansion to excited baryons \cite{Cohen:2003fv}.

In the context of our approach, where the baryon is treated as a system of $N_c$ 
quarks irrespective of its spin-flavor symmetry, the SU(3) breaking operators are defined as 
\begin{equation}\label{operatorB1}
B_1 = n_s,
\end{equation}
where $n_s$ is the number of strange quarks and  
\begin{equation}\label{operatorB2}
B_2 = \frac{1}{N_c}(L^i G^{i8} - \frac{1}{2 \sqrt{3}} L \cdot S),
\end{equation}
\begin{equation}\label{operatorB3}
B_3 =\frac{1}{N_c}(S^iG^{i8} - \frac{1}{2 \sqrt{3}} S\cdot S),
\end{equation}
where the angular momentum operator $L^i$, the spin operator $S^i$ and the component 8 of the 
spin-flavor operator $G^{i8}$ act on the entire system of  $N_c$ quarks.


\section{Matrix elements}\label{matrixelements}

\subsection{The multiplet $[{\bf 56},2^+]$}

\begin{table}
\caption{Matrix elements of $O_i$ for SU(3) octets and decuplets belonging to the $[56,2^+]$ multiplet.}
\label{56,2+}
\[
\renewcommand{\arraystretch}{1.5}
\begin{array}{crrrr}
\hline
\hline
           & \ \ \ \ \ \ \ \  O_1  & \ \ \ \ \ \ \ O_2  & \ \ \ \ \ \ \ O_3   & \ \ \ \ \ \ \ O_6 \\
\hline
^28[56,2^+]{\frac{3}{2}}^+    & N_c  & - \fr{3}{2 N_c} & \fr{3}{4 N_c}   & 0 \\
^28[56,2^+]{\frac{5}{2}}^+    & N_c  &   \fr{1}{  N_c} & \fr{3}{4 N_c}   & 0 \\
^410[56,2^+]{\frac{1}{2}}^+  & N_c  & - \fr{9}{2 N_c} & \fr{15}{4 N_c}   & \fr{7}{2 N_c} \\
^410[56,2^+]{\frac{3}{2}}^+  & N_c  & - \fr{3}{  N_c} & \fr{15}{4 N_c}   & 0 \\

^410[56,2^+]{\frac{5}{2}}^+  & N_c  & - \fr{1}{2 N_c} & \fr{15}{4 N_c}   & -\fr{5}{2 N_c} \\
^410[56,2^+]{\frac{7}{2}}^+  & N_c  &   \fr{3}{  N_c} & \fr{15}{4 N_c}   & \fr{1}{N_c} \\[0.5ex]
\hline
\hline
\end{array}
\]
\end{table}

In Table \ref{56,2+} we reproduce 
the analytic forms obtained for the operators $O_i$ in terms of $N_c$ for the multiplet $[{\bf 56},2^+]$.
The first  
is the trivial spin-flavor singlet operator $O_1$ defined by Eq. (\ref{massoperator}) with matrix elements equal to $N_c$ in all cases.
For symmetric states the spin-orbit operator $O_2$ and the spin operator $O_3$ 
are of order $\mathcal{O}(N^{-1}_c)$.
From Eq. (\ref{LS}) it follows that the matrix elements of $O_2$ are given by the usual formula
\begin{equation}\label{spinorbit}
\langle O_2 \rangle = \frac{1}{2 N_c} [J(J+1) - L(L+1) - S(S+1)],
\end{equation}
where $L$ is the angular momentum of the whole system.
The matrix elements of the spin operator $O_3$ are trivially equal to $\frac{1}{N_c}S(S+1)$. 
We are reminded that for symmetric states the matrix elements of the isospin operator $O_4$ are equal to those
of the spin operator, thus are not included in Table \ref{56,2+}.

We also include the operator $O_6$, defined 
by Eq. (\ref{lop}). In the $[{\bf 56},2^+]$ multiplet it contributes only to the decuplet resonances. 
The general analytic form of the matrix elements of $O_6$ were derived in Ref. \cite{Matagne:2011fr} and for convenience 
the formula is reproduced
in  Appendix  \ref{operatorO6}. The calculations require knowledge of the matrix elements of the SU(6) generators 
for spin-flavor symmetric states, namely the
isoscalar factors $[N_c] \times [21^4] \rightarrow [N_c]$,
derived in Table I of Ref. \cite{Matagne:2006xx}.

Although  $G^{ia}$ acts sometimes as a coherent operator, introducing an extra power of $N_c$ in mixed symmetric
states, this is not the case for spin-symmetric  states, so that the matrix elements of  $O_6$ are of order $1/N_c$,
as one can see from Table \ref{56,2+}.

\begin{table}
\caption{Expectation values of SU(3) breaking operators for strange octets of the $[56,2^+]$ multiplet.
 Here, $a_J = 1,-2/3$ for $J=3/2, 5/2$ respectively, from Ref. \cite{GSS03}.
}
\label{break56}
\[
\begin{array}{cccc}
\hline
\hline
  &  \hspace{ .6 cm}  {B}_1 \hspace{ .6 cm}  &
     \hspace{ .6 cm}  {B}_2 \hspace{ .6 cm}  &
     \hspace{ .6 cm}  {B}_3 \hspace{ .6 cm}   \\
\hline
^2\Lambda_{J} & 1 &   \fr{ 3 \sqrt{3} a_J}{4 N_c} &   
- \fr{3 \sqrt{3}}{8 N_c}   \\
^2\Sigma_{J}  & 1 &   - \fr{  \sqrt{3}\ a_J}{4 N_c} &     
\fr{  \sqrt{3}}{8 N_c}   \\
^2\Xi_{J}     & 2 &   \fr{  \sqrt{3}\ a_J}{N_c}   &   
- \fr{  \sqrt{3}}{2 N_c}   \\ [0.9ex]
\hline
\hline
\end{array}
\]
\end{table}

In the mass formula (\ref{massoperator}) we have included three first-order SU(3) symmetry 
breaking operators $B_1$, $B_2$ and $B_3$ 
(denoted by   $\bar {B}_1$, $\bar {B}_2$ and $\bar {B}_3$  in  Ref. \cite{GSS03}). 
Their nonvanishing matrix elements,
were calculated in Ref. \cite{GSS03} where the matrix element of the first term of $B_3$ 
of Eq. (\ref{operatorB3}) was obtained from the formula
\begin{equation}
\label{SGsym}
\langle S^i G^{i8} \rangle = \frac{1}{4 \sqrt{3}} [ 3I(I+1) - S(S+1) - \frac{3}{4} n_s(n_s+2)],
\end{equation}
later proven for symmetric states in Ref. \cite{Matagne:2006xx} where $n_s$ is the number of strange quarks.

For octets the nonvanishing expectation values of $B_i$  are reproduced in Table \ref{break56} which shows  
that the effect of $ B_2 $ depends on $J$.
At $J$ = 3/2 it increases the mass of  $\Lambda$ and lowers the mass of  $\Sigma$ while 
for $J$ = 5/2 it does the other way round.
 $B_3$  has the role of lowering the mass of $\Lambda$ while increasing 
the mass of $\Sigma$, irrespective of $J$. In all $B_2$ and $B_3$ remove the degeneracy due to $B_1 = 1$.  

For the $^410$ strange members of the $[{\bf 56},2^+]$ multiplet
the expectation values found in  Ref. \cite{GSS03} for $ B_2 $  and  $B_3$ 
can be written in a compact form, as one can see in Appendix \ref{BREAK}. These are
\begin{equation}\label{BREAK2}
 B_2  = - \frac{n_s}{2\sqrt{3} N_c} \langle L \cdot S \rangle,
\end{equation}
and
\begin{equation}\label{BREAK3}
 B_3  = - \frac{n_s}{2\sqrt{3} N_c} \langle S \cdot S \rangle.
\end{equation}
Eqs. (\ref{BREAK2}) and (\ref{BREAK3}) give equal space splitting between the decuplet members at fixed $J$.
Note that the operator $B_2$ can raise or lower the mass depending on the sign of $\langle L \cdot S \rangle$.
From Eqs. (\ref{BREAK2}) and (\ref{BREAK3}) and the definitions of $O_2$ and $O_3$ it follows that
the expectation values of $O_2$, $O_3$, $B_2$ and $B_3$ satisfy the relation
\begin{equation}\label{RATIO}
\frac{B_2}{B_3} = \frac{O_2}{O_3},
\end{equation}
which holds in fact for both the octet and the decuplet of the $[{\bf 56},2^+]$ multiplet \cite{Matagne:2004pm}.


\subsection{The multiplet $[\bf 70,\ell^+]$}

In Table \ref{BARYON70} we reproduce the analytic forms of the matrix elements of the operators $O_i$
included in the mass formula for the multiplet $[\bf 70,\ell^+]$. They are successively listed for all possible octets,
decuplets and SU(3)-flavor singlets of this multiplet. Details of the derivation of these 
analytic forms as a function of $N_c$ can be found in Ref. \cite{Matagne:2013cca}.
Accordingly, the spin-orbit operator $O_2$ is the single-particle operator 
\begin{equation}\label{newspinorbit}
O_2 = \ell \cdot s = \sum^{N_c}_{i=1} \ell(i) \cdot s(i),
\end{equation}
the matrix elements of which are of order $N^0_c$ and 
are calculated in Appendix \ref{operatorO2}.

Note that in the case of mixed symmetric states  the matrix elements of the operator $O_6$ are  $\mathcal{O}(N^0_c)$,
in contrast to the symmetric case where they are $\mathcal{O}(N^{-1}_c)$, and  nonvanishing only
for octets, while for the symmetric case they are nonvanishing for decuplets. 
Thus, at large $N_c$ the splitting starts at order $\mathcal{O}(N^{0}_c)$ for mixed symmetric states 
due both to $O_2$ and $O_6$,
while in the symmetric core + excited quark procedure \cite{CCGL} several operators are order $\mathcal{O}(N^{0}_c)$ \cite{Cohen:2005ct}. 
 
\begin{table}[htb]
\caption{Matrix elements of  $O_i$  operators for SU(3)-flavor octet, decuplet and singlet states in $[{\bf 70},\ell^+]$ multiplet.}
\label{BARYON70}
\renewcommand{\arraystretch}{1.2}
\begin{tabular}{lcccccc}
\hline
   &  \hspace{ .3 cm} $O_1$ \hspace{ .3 cm}  & \hspace{ .3 cm} $O_2$  \hspace{ .3 cm} & \hspace{ .3 cm} $O_3$  \hspace{ .3 cm}&  \hspace{ .3 cm} $O_4$  \hspace{ .3 cm}  
 & \hspace{ .3 cm} $O_6$ \\
\hline
$^48[{\bf 70},2^+]\frac{7}{2}^+$  &  $N_c$   &  $ 1 $    & $\frac{15}{4N_c}$ & $\frac{3}{4N_c}$ 
&  $-\frac{N_c-1}{4N_c}$\\
$^28[{\bf 70},2^+]\frac{5}{2}^+$  &  $N_c$   &  $\frac{(2N_c-3)}{3N_c}$    & $\frac{3}{4N_c}$  &   $\frac{3}{4N_c}$ 
& 0 \\
$^48[{\bf 70},2^+]\frac{5}{2}^+$  &  $N_c$   &  $-\frac{1}{6}$  & $\frac{15}{4N_c}$ &   $\frac{3}{4N_c}$ 
& $\frac{5(N_c-1)}{8N_c}$ \\
$^48[{\bf 70},0^+]\frac{3}{2}^+$  &  $N_c$   &  0    & $\frac{15}{4N_c}$ &   $\frac{3}{4N_c}$  
& 0  \\
$^28[{\bf 70},2^+]\frac{3}{2}^+$  &  $N_c$   &  $-\frac{2N_c-3}{2N_c}$   & $\frac{3}{4N_c}$  &  $\frac{3}{4N_c}$ 
& 0 \\
$^48[{\bf 70},2^+]\frac{3}{2}^+$  &  $N_c$   &   $- 1$   & $\frac{15}{4N_c}$ &  $\frac{3}{4N_c}$ 
& 0 \\
$^28[{\bf 70},0^+]\frac{1}{2}^+$  &  $N_c$   &   0   &  $\frac{3}{4N_c}$  & $\frac{3}{4N_c}$  
& 0 \\
$^48[{\bf 70},2^+]\frac{1}{2}^+$  &  $N_c$   &  - $\frac{3}{2}$    & $\frac{15}{4N_c}$ &  $\frac{3}{4N_c}$ 
&  $-\frac{7(N_c-1)}{8N_c}$ 
\vspace{0.2cm}\\
\hline
$^210[{\bf 70},2^+]\frac{5}{2}^+$  &   $N_c$   &- $\frac{1}{3}$ & $\frac{3}{4N_c}$  & $\frac{15}{4N_c}$    
& 0\\
$^210[{\bf 70},2^+]\frac{3}{2}^+$  &   $N_c$   &  $\frac{1}{2}$  & $\frac{3}{4N_c}$  & $\frac{15}{4N_c}$ 
& 0 \\
$^210[{\bf 70},0^+]\frac{1}{2}^+$  &   $N_c$   &  0              & $\frac{3}{4N_c}$ & $\frac{15}{4N_c}$ 
& 0 \vspace{0.2cm}\\
\hline
$^21[{\bf 70},2^+]\frac{5}{2}^+$  &   $N_c$   &  $ 1           $ & $\frac{3}{4N_c}$  & $-\frac{2 N_c + 3}{4N_c}$    
& 0\\
$^21[{\bf 70},2^+]\frac{3}{2}^+$  &   $N_c$   &  $ -\frac{3}{2}$ & $\frac{3}{4N_c}$  & $-\frac{2 N_c + 3}{4N_c}$ 
& 0 \\
$^21[{\bf 70},0^+]\frac{1}{2}^+$  &   $N_c$   &  0             & $\frac{3}{4N_c}$ & $-\frac{2 N_c + 3}{4N_c}$ 
& 0 \\[0.5ex]
\hline
\hline
\end{tabular}
\end{table}

\begin{table}
\caption{Matrix elements of the SU(3) breaking operators $B_i$ for strange baryons for
the $^48_{7/2}$, $^48_{5/2}$, $^28_{3/2}$, $^210_{5/2}$, $^210_{1/2}$  sectors and
the singlet $ ^2\Lambda^{'}_{1/2}$  of the $[{\bf 70},\ell^+]$ multiplet.}
\label{break70}
\[
\renewcommand{\arraystretch}{1.5}
\begin{array}{cccc}
\hline
\hline
  &  \hspace{ .6 cm}  B_1 \hspace{ .6 cm}  &
     \hspace{ .6 cm}   B_2  \hspace{ .6 cm}  &
     \hspace{ .6 cm}  {B}_3 \hspace{ .6 cm} \\
\hline
^4\Lambda_{7/2}  &  1  & - \fr{\sqrt{3}}{2N_c}               & - \fr{5\sqrt{3}}{8 N_c}  \\
^4\Sigma_{7/2}   &  1  &   \fr{\sqrt{3}}{6 N_c} \fr{N_c-9}{N_c-1}  &   \fr{ 5 \sqrt{3}}{24 N_c}  \fr{N_c-9}{N_c-1}  \\
^4\Xi_{7/2}      &  2  & - \fr{2\sqrt{3}}{3}\frac{1}{N_c-1}     &   - \fr{5\sqrt{3}}{6}\frac{1}{N_c-1}  \\ [0.8ex]
\hline
^4\Lambda_{5/2}  &  1  &   \fr{\sqrt{3}}{12N_c}                & - \fr{5\sqrt{3}}{8 N_c} 
\\
^4\Sigma_{5/2}   &  1  & - \fr{\sqrt{3}}{36 N_c} \fr{N_c-9}{N_c-1}  &   \fr{ 5 \sqrt{3}}{24 N_c}  \fr{N_c-9}{N_c-1}  \\
^4\Xi_{5/2}      &  2  &   \fr{\sqrt{3}}{9}\frac{1}{N_c-1}     &   - \fr{5\sqrt{3}}{6}\frac{1}{N_c-1}  \\ [0.8ex]
\hline
^2\Lambda_{3/2}   &  1  & - \fr{\sqrt{3}}{4 N_c}  \fr{N_c-9}{N_c+3}          &  \fr{\sqrt{3}}{8 N_c}  \fr{N_c-9}{N_c+3}  \\
^2\Sigma_{3/2}    &  1  &   \fr{\sqrt{3}}{12 N_c}  \fr{N_c+3}{N_c-1}         & - \fr{\sqrt{3}}{24 N_c} \fr{N_c+3}{N_c-1} \\
^2\Xi_{3/2}  &  2 & - \fr{\sqrt{3}}{3 N_c} \fr{N^2_c-12N_c+9}{(N_c-1)(N_c+3)}& \fr{  \sqrt{3}}{6 N_c} \fr{N^2_c-12N_c+9}{(N_c-1)(N_c+3)}\\ [0.8ex]
\hline
^2\Sigma^{\ast}_{5/2}  &  1  &  -\fr{\sqrt{3}}{18 N_c} \fr{5N_c+9}{N_c+5}  & - \fr{ \sqrt{3}}{24 N_c} \fr{5N_c+9}{N_c+5}  \\
^2\Xi^{\ast}_{5/2}     &  2  &  -\fr{ \sqrt{3}}{9 N_c} \fr{5N_c+9}{N_c+5}  & - \fr{ \sqrt{3}}{12 N_c} \fr{5N_c+9}{N_c+5} \\
^2\Omega_{5/2}         &  3  &  -\fr{ \sqrt{3}}{6 N_c} \fr{5N_c+9}{N_c+5}  & - \fr{ \sqrt{3}}{8 N_c} \fr{5N_c+9}{N_c+5}  \\[0.8ex]
\hline
^2\Sigma^{\ast}_{1/2}  &  1  &  0  & - \fr{ \sqrt{3}}{24 N_c} \fr{5N_c+9}{N_c+5}  \\
^2\Xi^{\ast}_{1/2}     &  2  &  0  & - \fr{ \sqrt{3}}{12 N_c} \fr{5N_c+9}{N_c+5} \\
^2\Omega_{1/2}         &  3  &  0  & - \fr{ \sqrt{3}}{8 N_c} \fr{5N_c+9}{N_c+5}  \\[0.8ex]
\hline
^2\Lambda^{'}_{1/2}        &  1  &                     0                               & - \fr{3\sqrt{3}}{8 N_c} \fr{N_c-1}{N_c+3} \\[0.8ex]
\hline
\hline
\end{array}
\]
\end{table}
The SU(3) flavor breaking operators $B_i$ are the same as for the symmetric multiplet. 
The  operator $B_1$ is given by Eq. (\ref{operatorB1}). 
The matrix  elements of $B_2$ and $B_3$ were calculated as explained in Appendix \ref{BREAK}.
Results for octets and decuplets at some fixed $J$ are exhibited in Table  \ref{break70}.

For practical purposes we have summarized
these results by two simple analytic formulas valid at $N_c$ = 3. 
The diagonal matrix elements of $B_2$ take the following form
\begin{equation}\label{B2}
  B_2 = - n_s 
\frac{\langle L \cdot S \rangle}{6 \sqrt{3}},
\end{equation}
where ${\langle L \cdot S \rangle}$ is the expectation value of the spin-orbit operator with
the angular momentum operator acting on the whole system. 
Similarly 
the diagonal matrix elements of $B_3$ take the simple analytic form 
\begin{equation}\label{B3}
  B_3  = - n_s
\frac{S(S + 1)}{6 \sqrt{3}},
\end{equation}
where  $S$ is the total spin. The contribution of $B_3$ is always negative, otherwise vanishing 
for nonstrange baryons.
These formulas can be applied to  $^28_J$, $^48_J$, 
$^2{10}_J$ and $^2{1}_{1/2}$ baryons of the  $[{\bf 70},\ell^+]$ multiplet. From Eqs. (\ref{B2}) and (\ref{B3})
it follows that Eq. (\ref{RATIO}) is satisfied for the $[{\bf 70},2+]$ multiplet as well.

\section{Fit and discussion}\label{fit}

Presently we perform a consistent  analysis of the experimentally known 
resonances supposed to belong either to the symmetric  
$[{\bf 56},2^+]$ multiplet or 
to the mixed symmetric multiplet $[{\bf 70},\ell^+]$ with $\ell$ = 0 or 2,
by using the same operator basis. Results of the fitted coefficients $c_i$ and
$d_i$ are exhibited in  Table \ref{operators} together with the values of $\chi_{\mathrm{dof}}^2$
for each multiplet.

For the $[{\bf 56},2^+]$ multiplet the values of the coefficients $c_i$ and $d_i$ are quite close 
to those of  Ref. \cite{GSS03}. The spin-orbit coefficient $c_2$ is about twice as small in the present
case but one should  take into account the contribution of the operator  $O_6$, 
also depending on the angular momentum, which is absent in Ref. \cite{GSS03}. The PDG2014 data
give a somewhat larger $\chi_{\mathrm{dof}}^2$ as compared to the PDG2002 data used in Ref. \cite{GSS03}
where $\chi_{\mathrm{dof}}^2$ was reported to be 0.7. But this does not much affect the coefficients 
$c_i$ or $d_i$.  In the fit for the $[{\bf 56},2^+]$ multiplet we have ignored the poorly known resonance $\Sigma(1840)3/2^{+*}$ 
by analogy with Ref. \cite{GSS03}.

The results for the $[{\bf 70},\ell^+]$ multiplet can only roughly be compared to those Table I, Fit 2
of Ref. \cite{Matagne:2013cca}, because  $B_2$ and $B_3$ were missing there. Note that the
factor 15 of $O_6$ has been removed here, which explains the larger value of $c_6$ now.
In fact the product $c_6 O_6$ matters in the mass. The values of $c_2$  are similar to Ref. \cite{Matagne:2013cca}. 
The $1/N_c$ corrections
are dominated by $O_3$ and $O_4$. The sum of $c_3$ and $c_4$ is comparable to that of $c_3$
in $[56,2^+]$ where $O_3$ and $O_4$ contribute equally.

An important remark is that the values of the most dominant coefficient $c_1$
are different for the two multiplets. The $c_1$ of $[{\bf 70},\ell^+]$ is higher by about 85 MeV.
This implies that two distinct
Regge trajectories are expected for the symmetric and mixed symmetric multiplets 
as already hinted in Ref. \cite{Matagne:2013cca}.

As mentioned in the Introduction, in quark models $c_1^2$ would correspond to the kinetic plus the confinement 
parts of the spin-independent Hamiltonian \cite{Matagne:2014lla,Semay:2007cv}. In a semirelativistic 
model it happens that the multiplet $[{\bf 70},\ell^+]$ lies above $[{\bf 56},\ell^+]$ when the
hyperfine interaction is switched off.  For example, in Ref. \cite{Stassart:1997vk}
it was explicitly shown that the $[{\bf 70},4^+]$ multiplet is
about 50 MeV higher than the $[{\bf 56},4^+]$ multiplet. Therefore the present results for $c_1$ hint 
at a qualitative agreement with the quark model.

\begin{table}[htb]
\caption{List of dominant operators and their coefficients (MeV) from the mass formula (\ref{massoperator}) obtained 
in numerical fits for the $[{\bf 56},2^+]$ in column 2 and $[{\bf 70},\ell^+]$ multiplets in columns 3 and 4 respectively.
The spin-orbit operator $O_2$ is defined by Eq. (\ref{LS}) for $[{\bf 56},2^+]$ and by Eq.(\ref{newspinorbit})
for $[{\bf 70},\ell^+]$. }
\label{operators}
\renewcommand{\arraystretch}{1.2} 
\begin{tabular}{lrrrrr}
\hline
\hline
Operator & $[{\bf 56},2^+]$ &\hspace{0.5cm} &  $[{\bf 70},\ell^+]$ \hspace{0.01cm}  &\\
\hline
\hline
$O_1 = N_c \ \1 $                               &   542 $\pm$ 2    & &   627 $\pm$ 10 &    \\
$O_2$ spin-orbit                  &   Eq.(10)   7 $\pm$10   & & Eq.(15)  69 $\pm$ 26 &    \\
$O_3 = \frac{1}{N_c}S^iS^i$                     &   233 $\pm$ 11   & &    88 $\pm$ 31 &     \\
$O_4 = \frac{1}{N_c}\left[T^aT^a-\frac{1}{12}N_c(N_c+6)\right]$ &  & &   127 $\pm$ 21 &   \\[0.8ex]
$O_6 =  \frac{1}{N_c} L^{(2)ij} G^{ia} G^{ja}$ &   6 $\pm 19$      & &    72 $\pm$ 71 &    \\[0.5ex]
\hline
$B_1 = n_s$                                     &  205 $\pm$ 14    & &     76 $\pm$ 31 &      \\ 
$B_2 = \frac{1}{N_c}(L^iG^{i8}  - \frac{1}{2\sqrt{3}} L^i S^i)$    &    97 $\pm$ 40     & & - 172 $\pm$ 106 & \\
$B_3 = \frac{1}{N_c}(S^iG^{i8}  - \frac{1}{2\sqrt{3}} S^i S^i)$    & 197 $\pm$ 69       & &    279 $\pm$ 144 &  \\[0.8ex]
\hline                  
$\chi_{\mathrm{dof}}^2$                                            &     1.63           & &    1.48 &    \\
\hline \hline
\end{tabular}
\end{table}

\subsection{The multiplet $[{\bf 56},2^+]$}

The partial contribution and the calculated total mass obtained from the fit are presented in Table \ref{symm}.
The experimental masses are taken from the 2014 version of the Review of Particle Properties (PDG) \cite{PDG},
except for $\Delta(1905)5/2^+$ where we used the mass of Ref. \cite{GSS03} which gives a smaller $\chi_{\mathrm{dof}}^2$,
but does not much change the fitted values of $c_i$ and $d_i$. As expected, the most important subleading contribution comes 
from the spin operator $O_3$. The contributions of the angular momentum-dependent operators $O_2$ and $O_6$ are 
comparable, but small. Among the SU(3) breaking terms, $B_1$ is dominant. An important remark is that 
in the  $[{\bf 56},2^+]$ multiplet $B_2$ and $B_3$ lift the degeneracy of $\Lambda$ and $\Sigma$ baryons in 
the octets, which is not the case for the $[{\bf 70},\ell^+]$ multiplet.


{\squeezetable
\begin{table}
\caption{Partial contribution and the total mass (MeV) predicted by the $1/N_c$ expansion  
with operators of Tables \ref{56,2+} and \ref{break56}.
The last two columns give  the empirically known masses and the 2014 status in the Review of Particles Properties 
\cite{PDG}.}
\label{symm}
\begin{tabular}{crrrrrrcccccccc}\hline \hline
                    &      \multicolumn{7}{c}{Partial contrib. (MeV)} & \hspace{.0cm} Total (MeV)  \hspace{.0cm}  &
                                              \hspace{.0cm}   Experiment(MeV)   \hspace{.0cm}  &  Name, Status\\
\cline{2-8}
                    &      \hspace{.0cm} $c_1O_1$  & \hspace{.0cm}  $c_2O_2$ &
                            \hspace{.0cm} $c_3O_3$  & \hspace{.0cm} $c_6O_6$  &
                           \hspace{.0cm} $d_1B_1$ & \hspace{.0cm} $d_2B_2$ & \hspace{.0cm} $d_3B_3$ 
                                                                                                & \hspace{0cm}  &  & \\
\hline
$^28N[56,2^+]{\frac{3}{2}}^+$        &  1626  & - 4  &  58  &  0 &   0    &   0   &  0  &   $ 1680\pm9 $  & $1725\pm25$ 
& $N(1720){\frac{3}{2}}^{+}$****\\
$^28\Lambda[56,2^+]{\frac{3}{2}}^+$  &        &      &      &    &  205   &  42   &- 42  &  $ 1885\pm29 $ & $1880\pm30$ 
& $\Lambda(1890){\frac{3}{2}}^{+}$****\\
$^28\Sigma[56,2^+]{\frac{3}{2}}^+$   &        &      &      &    &  205   & -14   &  14  &  $ 1885\pm18 $   & 
\\
$^28\Xi[56,2^+]{\frac{3}{2}}^+$      &        &      &      &    &  410   & -56   &  57  &  $ 2089\pm40 $       \vspace{0.1cm}    \\
\hline
$^28N[56,2^+]{\frac{5}{2}}^+$        &  1626  &  2     &  58 &  0 &   0    &   0   &  0  &  $1686\pm5 $  &  $ 1685\pm5 $ 
& $N(1680){\frac{5}{2}}^{+}$****\\
$^28\Lambda[56,2^+]{\frac{5}{2}}^+$ &	      &        &     &    & 205    & -28   & -42 &  $1821\pm5 $  & $ 1820\pm5 $
 & $\Lambda(1820){\frac{5}{2}}^{+}$****\\
$^28\Sigma[56,2^+]{\frac{5}{2}}^+$  &         &        &     &    & 205    &  9    &  14 &  $1915\pm17 $  & $ 1918\pm18$ 
& $\Sigma(1915){\frac{5}{2}}^{+}$****\\
$^28\Xi[56,2^+]{\frac{5}{2}}^+$     &         &        &     &    & 410    & -37   & -57 &  $2002\pm13 $  & \vspace{0.1cm} \\
\hline
$^410\Delta[56,2^+]{\frac{1}{2}}^+$  &  1626    &  - 10   & 291  &  7   &   0   &   0   &  0   &  $ 1914\pm27$  & $ 1890\pm30 $
& $\Delta(1910){\frac{1}{2}}^{+}$****\\
$^410\Sigma[56,2^+]{\frac{1}{2}}^{+}$&          &         &      &      &  205  &  28   & -71  &  $ 2076\pm35$  &                \\
$^410\Xi[56,2^+]{\frac{1}{2}}^{+}$   &          &         &      &      &  410  &  56   & -142 &  $ 2238\pm58$  &                \\
$^410\Omega[56,2^+]{\frac{1}{2}}^{+}$&          &         &      &      &  615  &  84   & -213 &  $ 2400\pm85$  &   \vspace{0.1cm}\\
\hline
$^410\Delta[56,2^+]{\frac{3}{2}}^+$ &  1626     &  - 7    &  291 &   0  &   0   &   0   &  0   &  $1910\pm18$  &  $1935\pm35$ 
& $\Delta(1920){\frac{3}{2}}^{+}$*** \\
$^410\Sigma[56,2^+]{\frac{3}{2}}^+$ &           &         &      &      &  205  &  28   & -71  &  $2072\pm31$  & 
& \\
$^410\Xi[56,2^+]{\frac{3}{2}}^+$    &           &         &      &      &  410  &  56   & -142 &  $2234\pm57$  &            \\
$^410\Omega[56,2^+]{\frac{3}{2}}^+$ &           &         &      &      &  615  &  84   & -213 &  $2396\pm85$  &      \vspace{0.1cm}\\
\hline
$^410\Delta[56,2^+]{\frac{5}{2}}^+$ &  1626    &   - 1    & 291  & - 5  &   0   &   0   &  0   &  $1911\pm23$  & $1895 \pm25$  
& $\Delta(1905){\frac{5}{2}}^{+}$****\\
$^410\Sigma[56,2^+]{\frac{5}{2}}^+$ &          &          &      &      &  205  &   5   &  -71 &  $2050\pm29$  & 
\\
$^410\Xi[56,2^+]{\frac{5}{2}}^+$    &          &          &      &      &  410  &   9   & -142 &  $2188\pm44$  &                 \\
$^410\Omega[56,2^+]{\frac{5}{2}}^+$&          &          &      &      &  615  &  14    & -213 &  $2326\pm63$  &    \vspace{0.1cm}  \\
\hline
$^410\Delta[56,2^+]{\frac{7}{2}}^+$ &   1626   &    7    & 291   &  2    &   0   &   0   &   0  &  $1926\pm12$ & $ 1930\pm20$   
& $\Delta(1950){\frac{7}{2}}^{+}$****\\
$^410\Sigma[56,2^+]{\frac{7}{2}}^+$ &          &         &       &       & 205  & -28    &  -71 &  $2032\pm8$ & $ 2033\pm 8$ 
& $\Sigma(2030){\frac{7}{2}}^{+}$****\\
$^410\Xi[56,2^+]{\frac{7}{2}}^+$    &          &          &      &       & 410  & -56    & -142 &  $2138\pm15 $  &       \\
$^410\Omega[56,2^+]{\frac{7}{2}}^+$ &          &          &      &       & 615  & -84    & -213 &  $2244\pm26 $  & \vspace{0.1cm}  \\
\hline \hline
\end{tabular}
\end{table}}


\subsection{The multiplet $[{\bf 70},\ell^+]$}

The 2014 version of the Review of Particle Properties (PDG) \cite{PDG}  incorporates the new 
multichannel partial wave analysis of the Bonn-Gatchina group  \cite{Anisovich:2011fc}. 
According to the Bonn-Gatchina group
the resonance $P_{13}(1900)$ has been upgraded from two to three stars with a Breit-Wigner mass of
1905 $\pm$  30 MeV.
The resonance  $N(2000)5/2^+$ 
has been split into two two-star resonances, namely $N(1860)5/2^+$ and $N(2000)5/2^+$,
with masses  indicated in Table \ref{MASSES}.
There is a new one-star resonance $N(2040)3/2^+$ observed in the decay $J/\psi \rightarrow p \bar p \pi^0$
\cite{Ablikim:2009iw}.
There is also a new two-star resonance $N(1880)1/2^+$  observed by the Bonn-Gatchina group with a mass of
1870 $\pm$ 35 MeV \cite{Anisovich:2011fc}.

As compared to Ref. \cite{Matagne:2013cca} where only 11 resonances have been included in the numerical fit, here
we consider 16 resonances, having a status of three, two or one star.
This means that we have tentatively added  the resonances
$\Xi(2120)?^{?*}$, $\Sigma(2070)5/2^{+*}$, $\Sigma(1940)?^{?*}$,  $\Xi(1950)?^{?***}$ and
$\Sigma(2080)3/2^{+**}$. The masses and the error bars considered
in the fit correspond to averages over those data taken into account in the particle listings,
except for a few which favor specific experimental values cited in the headings of Table \ref{MASSES}.
For example the value of the mass of the $N(1880)1/2^{+**}$ resonance is taken identical to
that of Ref. \cite{Anisovich:2011fc}, the other data mentioned in the listings being ignored.
For $\Delta(2000)5/2^{+**}$ and $\Sigma(1880)1/2^{+**}$ we averaged over two and eight experimental 
values, respectively, indicated in the 2014 version of PDG.

We have ignored the $N(1710){1/2^{+***}}$ and the 
$\Sigma(1770){1/2^{+*}}$ resonances, 
the theoretical argument being that their masses are too low, leading to unnatural sizes for the 
coefficients  $c_i$ or $d_i$ \cite{Pirjol:2003ye}.
On the experimental side one can justify the removal of  the $N(1710)1/2^{+***}$ resonance due to
the latest GWU analysis  of Arndt et al. \cite{Arndt:2006bf} where it has not been seen. 
This is a controversial resonance.
We have also ignored
the $\Delta(1750)1/2^{+*}$ resonance, considered previously \cite{Matagne:2006zf}, inasmuch as,
neither Arndt et al. \cite{Arndt:2006bf} nor Anisovich et al.  \cite{Anisovich:2011fc} find evidence
for it.

{\squeezetable
\begin{table}
\caption{Partial contribution and the total mass (MeV) predicted by the $1/N_c$ expansion 
with operators of Tables \ref{BARYON70} and \ref{break70}.
The last two columns give  the empirically known masses and status from the 2014 Review of Particles Properties 
\cite{PDG} unless specified by (A) from \cite{Anisovich:2011fc}, 
(L) from \cite{Litchfield:1971ri},
(Z) from \cite{Zhang:2013sva},
(G1) from  \cite{Gopal:1980ur}, (B) from \cite{Biagi:1986vs},
(AB) from \cite{Ablikim:2009iw},
(G2) from  \cite{Gopal:1976gs},
.}\label{MASSES}
\renewcommand{\arraystretch}{1.5}
\begin{tabular}{crrrrrrcccccl}\hline \hline
                    &      \multicolumn{8}{c}{Part. contrib. (MeV)}  & \hspace{0.0cm} Total (MeV)   & \hspace{.0cm}  Experiment (MeV)\hspace{0.0cm}
 &\hspace{0.cm}  Name, status \hspace{.0cm} \\

\cline{2-9}
                    & \hspace{.0cm} $c_1O_1$  & \hspace{.0cm}  $c_2O_2$ & \hspace{.0cm}$c_3O_3$ &\hspace{.0cm}  $c_4O_4$ 
&\hspace{.0cm}  $c_6O_6$ & $d_1B_1$ & $d_2B_2$ & $d_3B_3$ &  \\
\hline
$^4N[{\bf 70},2^+]\frac{7}{2}^+$        & 1882 & 69 & 109 & 32 & - 12 & 0  &  0  &    0  & $2080\pm 39$  & $2060\pm65$(A) & $N(1990)7/2^+$**  \\
$^4\Lambda[{\bf 70},2^+]\frac{7}{2}^+$  &      &    &     &    &      & 76 & 50  & -101  & $2105\pm 19$  & $2100\pm30$(L) 
& $\Lambda(2020)7/2^+$* \\
$^4\Xi[{\bf 70},2^+]\frac{7}{2}^+$      &      &    &     &    &    &  152 &  99 & - 201 & $2130\pm 8$ & 
$2130\pm8$  &  $\Xi(2120)?^?$* \vspace{0.2cm}\\
\hline
$^4N[{\bf 70},2^+]\frac{5}{2}^+$        & 1882 & - 12  & 109 & 32 & 30 & 0  & 0  & 0  & $2042\pm41$ & $2000\pm50$ & $N(2000)5/2^+$**
\vspace{0.2cm} \\
$^4\Lambda[{\bf 70},2^+]\frac{5}{2}^+$  &      &       &     &    &    & 76 & -8 & -101 & $2009\pm40$ &  &  \\
\hline
$^4N[{\bf 70},2^+]\frac{3}{2}^+$        & 1882 & -69   &  109 & 32  & 0  &  0  &  0  & 0     & $1955\pm32$       &    & \vspace{0.2cm} \\
\hline
$^4N[{\bf 70},2^+]\frac{1}{2}^+$        & 1882 &- 103 & 109 & 32 &- 42 & 0 & 0 & 0  & $1878\pm34 $ & $1870\pm35$(A) & $N(1880)1/2^+$**
\vspace{0.2cm}\\
 \hline
$^2N[{\bf 70},2^+]\frac{5}{2}^+$        & 1882 & 23  & 22  & 32  & 0   & 0   &  0 &  0   & $1959\pm29$ & $1860\pm{^{120}_{60}}$(A) & $N(1860)5/2^+$** \\
$^2\Lambda[{\bf 70},2^+]\frac{5}{2}^+$  &      &     &     &     & 0   & 76  & 17 & - 20 & $2031\pm11$ &  $2036\pm13$(Z)  
&  $\Lambda(2110)5/2^+$***           \\
$^2\Sigma[{\bf 70},2^+]\frac{5}{2}^+$   &      &     &     &     & 0   & 73  & 15 & - 19 & $2031\pm11$ &  $2051\pm25$(G1) 
& $\Sigma(2070)5/2^+$* \vspace{0.2cm}\\
\hline
$^2N[{\bf 70},2^+]\frac{3}{2}^+$        & 1882 &- 34 & 22  & 32  & 0   &  0  & 0  & 0   & $1902\pm22$ & $1905\pm30$(A) & $N(1900)3/2^+$***  \\
$^2\Sigma[{\bf 70},2^+]\frac{3}{2}^+$   &      &     &     &     & 0   &  76 & - 25 & - 20 & $1933\pm11$ & $1941\pm18$ & $\Sigma(1940)?^?$*  
\vspace{0.2cm} \\
$^2\Xi[{\bf 70},2^+]\frac{3}{2}^+$      &      &     &     &     & 0   & 152 & - 50 & - 40 & $1964\pm70$ & $1967\pm7$(B) & $\Xi(1950)?^?$*** \vspace{0.2cm}\\
\hline
$^4N[{\bf 70},0^+]\frac{3}{2}^+$         & 1882 &  0  & 109 & 32  &  0  &  0  & 0 & 0      & $2024\pm20$ & $2040\pm28$(AB)  & $N(2040)3/2^+$*\\
$^4\Sigma[{\bf 70},0^+]\frac{3}{2}$    &      &     &     &     &     &  76 & 0 & - 101  & $2000\pm23$&  $2100\pm69$   
& $\Sigma(2080)3/2^+$**\vspace{0.2cm} \\
\hline
$^2\Delta[{\bf 70},2^+]\frac{5}{2}^+$       & 1882  & 23  & 22 & 159  & 0  &  0 & 0 &  0  & $2086\pm37$  & $1962\pm139$ & 
$\Delta(2000)5/2^+$**\vspace{0.2cm}\\
\hline
$^2\Sigma^{\ast}[{\bf 70},0^+]\frac{1}{2}^+$& 1882  & 0   & 22 & 159  & 0  & 76 & 0 & -20 & $2119\pm25$   & $1902\pm96$  & $\Sigma(1880)1/2^+$** 
\vspace{0.2cm}\\
\hline
$^2\Lambda'[{\bf 70},0^+]\frac{1}{2}^+$     & 1882  & 0    & 22 & - 95 & 0  & 76 & 0 & - 20 & $1865\pm19$   & $1853\pm20$(G2) & $\Lambda(1810)1/2^+$*** 
\vspace{0.2cm}\\
\hline
\hline
\end{tabular}
\end{table}}

The partial contributions and the calculated total masses obtained from the fit are presented in Table  \ref{MASSES}.
One can see that the fit is good 
except for $\Sigma(1880)1/2^{+**}$ where the calculated mass is too high, perhaps suggesting that the 
average over the eight resonances indicated in the particle listings is not quite adequate.

Regarding the contribution of various operators we note that the good fit for $N(1880)1/2^{+**}$
was due to contribution of the spin-orbit operator $O_2$ of -$103$ MeV and of the operator 
$O_6$ which contributed with $-42$ MeV. The good fit also suggests that  $\Sigma(1940)?^{?*}$
and $\Xi(1950)?^{?***}$ assigned by us to the $^2[{\bf 70},2^+]3/2^+$ multiplet is well justified and that these 
resonances may have $J^P$ = $3/2^+$,  hopefully  to be confirmed experimentally in future analyses. 

The $1/N_c$ expansion is based on the SU(6) symmetry which naturally allows a classification of
excited baryons into octets, decuplets and singlets. In Table \ref{MASSES} 
the experimentally known resonances are presented. In addition some predictions 
are made for unknown resonances.    Many of the partners in a given
supermultiplet are not known. Note that $\Lambda$ and $\Sigma$ are degenerate in our approach.

The present findings can be compared to the suggestions for assignments
in the $[{\bf 70},\ell^+]$ multiplet made in Ref. \cite{Crede:2013kia} as educated guesses. 
The assignment  of  $\Sigma(1880)1/2^{+**}$ 
as a $[{\bf 70},0^+]1/2^+$ decuplet resonance is  confirmed as well as the assignment  of $\Lambda(1810)1/2^{+***}$ 
as a flavor singlet.

However, we are at variance with Ref.  \cite{Crede:2013kia} regarding 
$\Lambda(2110)5/2^{+***}$ as a partner of $N(2000)5/2^{+**}$ in a spin quartet. 
Our suggestion is that $\Lambda(2110)5/2^{+***}$ is a member of a spin doublet, 
together with  $N(1860)5/2^{+**}$ and $\Sigma(2070)5/2^{+*}$.  
We also disagree that $N(1900)3/2^{+***}$ is a member of a spin quartet.
We propose it as a partner of $\Sigma(1940)?^{?*}$ and $\Xi(1950)?^{?***}$
in a spin doublet.

However, one has to keep in mind that at the same $J$ spin doublets
and quartets can mix, for example for $N[70,2^+]$ at $J$ = 3/2 or 5/2.
The mixing would be due to the off-diagonal matrix elements of the spin-orbit
operator $O_2$ and the tensor operator $O_6$. A qualitative simplified discussion is 
given in Appendix \ref{Mixing}. We plan further studies on this subject in the future.  

The problem of assignment is not trivial.
Within  the  $1/N_c$ expansion method Ref. \cite{GSS03} suggests that  
$\Sigma(2080)3/2^{+**}$ and $\Sigma(2070)5/2^{+*}$ could be members 
of two distinct decuplets in the $[{\bf 56},2^+]$ multiplet. It would be interesting 
to further investigate more hyperons hopefully based on more extended and reliable data. 

Note that the resonance $N(2040)3/2^{+*}$ is here identified as a member of a spin quartet in the 
$[{\bf 70},0^+]$ multiplet while Crede and Roberts  \cite{Crede:2013kia} interpret it as a member of
the SU(6) antisymmetric $[{\bf 20}, 1^+]$ multiplet, ignored so far, as not being physically possible.

Finally, we mention that although the operators $B_2$ and $B_3$ have different analytic forms 
at arbitrary $N_c$, as seen from Table  \ref{break70}, they acquire identical values at $N_c$ = 3
for $\Lambda$ and $\Sigma$ in octets, which means that they cannot  lift the degeneracy between these 
hyperons, as happens for the $[{\bf 56},2^+]$ multiplet. One can lift this degeneracy by introducing 
a new operator 
\begin{equation}
B_4 = \frac{1}{N_c} \sum_i^3 T^i T^i - O_4,
\end{equation}
as proposed in Ref. \cite{Matagne:2011fr}. Presently this is not necessary inasmuch as the experimental
data are too scarce and not  accurate enough. In addition, in some multiplets the hierarchy of masses 
as a function of the strangeness is contrary to expectations, for example for the 
multiplet $^4[70,2^+]5/2^+$.  This requires further investigation.


\section{Summary and conclusions}

The value obtained for the coefficient $c_1$ is about 85 MeV larger in the $[{\bf 70},\ell^+]$ multiplet than in the 
$[{\bf 56},2^+]$ multiplet. This implies that two distinct Regge-type trajectories are expected for the 
symmetric and mixed symmetric multiplets, consistent with previous literature \cite{Matagne:2013cca,Goity:2007sc}.
 The spin-orbit coefficient for the $[{\bf 70},\ell^+]$ multiplet, is similar to our previous 
work \cite{Matagne:2014lla}.
The spin and flavor operators are two-body and bring important contributions to the masses.
As one can see from  Table \ref{BARYON70}
the expectation values  of  $O_4$   are positive for octets and decuplets 
and of order  $N^{-1}_c$, as in SU(4), and negative  and of order $N^0_c$  for flavor singlets,
which makes its role rather subtle in the numerical fit, improving the singlet masses.  
The contribution of the operator $O_6$ containing an SO(3) tensor is important especially for  $[{\bf 70},\ell^+]$ multiplet.
Together with the spin-orbit it may lead  to the mixing of doublets and quartets to be considered in further  studies when the accuracy of 
data will increase. The incorporation of $B_2$ and $B_3$ in the mass formula of the $[{\bf 70},\ell^+]$ multiplet
brings more insight into  the SU(6) multiplet classification of excited baryons in the $N$ = 2 band.


\vspace{1cm}

\centerline{\bf Acknowledgments}

F.S. acknowledges support from the Fonds de la Recherche Scientifique - FNRS under the
Grant No. 4.4501.05.

\appendix

\section{Matrix elements of $O_2$ for the $[{\bf 70},\ell^+]$ multiplets }\label{operatorO2}

The expression of the matrix elements of the single-particle spin-orbit operator
of Eq.  (\ref{newspinorbit}) was first given in Ref. \cite{CCGL}.
In Ref. \cite{Cohen:2003fv} it was reproduced in a more concise form, written below 
in a slightly different form. Like for the operator $O_6$ given in the next 
appendix, every matrix element can be factorized into a part dependent on $\ell, S, S', J$
and another one, independent of   $\ell$ and $J$, but dependent on $N_c$. One has
 \begin{eqnarray}\label{O2}
\lefteqn{  \langle (\lambda'\mu')Y'I'I_3';\ell'S' JJ_3
 |\ell \cdot s|(\lambda\mu)YII_3;\ell SJJ_3\rangle  = } \nonumber \\ & & 
   \delta_{\ell'\ell}\delta_{\lambda\lambda'}\delta_{\mu\mu'} 
 \delta_{Y'Y} \delta_{I'I} \delta_{I_3'I_3}  
 C_{so}(\ell,S,S',J) F_{so}(N_c,S,S'),
 \end{eqnarray}  
where
 \begin{eqnarray}
C_{so}(\ell,S,S',J) & = &  (-1)^{J+\ell -I}
 \sqrt{\frac{3}{2}{(2S+1)(2S'+1)\ell(\ell+1)(2\ell+1)}} 
  \left\{\begin{array}{ccc}
   \ell & \ell & 1 \\
      S & S' & J 
  \end{array}\right\}.\nonumber \\
\end{eqnarray}
and 
\begin{eqnarray}
F_{so}(N_c,S,S')& = &
\sum_{\eta=\pm1}(-1)^{(1/2-\eta/2)}\left\{\begin{array}{ccc}
   1 & 1/2 & 1/2 \\
   S_c & S & S'
  \end{array}\right\} C_{\rho' \eta} C_{\rho \eta}
\end{eqnarray}
where $\rho = S-I$, $\eta = 1$ for $I_c = I+1/2$,  $\eta = - 1$ for $I_c = I- 1/2$ and
\begin{eqnarray}
C_{0+}  = \sqrt{\frac{S[N_c+2(S+1)]}{N_c(2S+1)}} \nonumber \\
C_{0-}  = - \sqrt{\frac{(S+1)(N_c-2S)}{N_c(2S+1)}},
\end{eqnarray}
for the $S = I$ nonstrange states.

Note that the matrix elements obtained from Eq. (\ref{O2}) have an extra factor 3/2 as 
compared to those derived in
Ref. \cite{Matagne:2006zf} for mixed symmetric multiplets. The reason  
is that in Ref. \cite{Matagne:2006zf} we have used an approximate orbital wave function
for simplicity. The second term of Eq. (5) of that reference was neglected as being of order
$N^{-1/2}_c$. Equation (\ref{O2}) contains the contribution of that part of the wave function.
Also note that Eq. (\ref{O2}) has been independently derived in Ref. \cite{Matagne:2012tm}
where the coefficients $c_{\rho \eta}$ were identified with isoscalar factors of the
permutation group $S_{N_c}$.

\section{Expectation values of $O_6$}\label{operatorO6}

The operator $O_6$, defined by Eq. (\ref{lop}), is proportional to 
$L^{(2)} \cdot G \cdot G$,  
where the SO(3) rank-two tensor $L^{(2)}$ is defined by Eq. (\ref{TENSOR}).
For a given $\ell$,  its matrix elements  can be rewritten in the following  
factorized form \cite{Matagne:2011fr}
 \begin{eqnarray}\label{O6}
\lefteqn{  \langle (\lambda'\mu')Y'I'I_3';\ell'S' JJ_3
 |(-1)^{i+j+a}L^{(2)ij}G^{-ia}G^{-j,-a}|(\lambda\mu)YII_3;\ell SJJ_3\rangle  = } \nonumber \\ & & 
   \delta_{\ell'\ell}\delta_{\lambda\lambda'}\delta_{\mu\mu'} 
 \delta_{Y'Y} \delta_{I'I} \delta_{I_3'I_3}  
 C(\ell,S,S',J) F(N_c,S,S'),
 \end{eqnarray} 
which multiplied by the factor $\frac{1}{N_c}$  gives the matrix elements of $O_6$.
The formula (\ref{O6}) contains a factor independent of $N_c$ which we have denoted by $C(\ell,S,S',J)$. This is 
 \begin{eqnarray}
C(\ell,S,S',J) & = & \delta_{S'S} (-1)^{J+\ell -S}\nonumber \\
& \times & \frac{1}{2} 
 \sqrt{\frac{5\ell(\ell+1)(2\ell-1)(2\ell+1)(2\ell+3)}{6}} \sqrt{(2S+1)(2S'+1)}
  \left\{\begin{array}{ccc}
   \ell & \ell & 2 \\
      S & S' & J 
  \end{array}\right\}.\nonumber \\
\end{eqnarray}
This factor can be used to calculate matrix elements of other symmetric multiplets than  $[56,2^+]$ by using the property
\begin{equation}
\frac{{\langle O_6 \rangle}_{[56,\ell^+]}}{{\langle O_6 \rangle}_{[56,2^+]}} =
\frac {C(\ell,S,S',J)} {C(2,S,S',J)}.
\end{equation}
The other factor of Eq. (\ref{O6}),  let us denote by  it $F(N_c,S,S')$,  is independent of $\ell$ but dependent on $N_c$ 
and of the representation $[f]$ of SU(6), containing the Casimir operator $C^{SU(6)}_{[N_c]}$.
This factor is 
\begin{eqnarray}
F(N_c,S,S')& = &C^{SU(6)}_{[f]}
\sum_{S''}(-1)^{(S-S'')}\left\{\begin{array}{ccc}
   1 & 1 & 2 \\
   S & S' & S''
  \end{array}\right\} 
  \nonumber \\& \times & \sum_{\rho,\lambda'',\mu''}
   \left(\begin{array}{cc||c}
         [f] & [21^4] & [f] \\
	 (\lambda''\mu'')S'' & (11)1 & (\lambda\mu)S 
        \end{array}\right)_{\rho} 
    \left(\begin{array}{cc||c}
         [f] & [21^4] & [f] \\
	 (\lambda''\mu'')S'' & (11)1 & (\lambda\mu)S' 
        \end{array}\right)_{\rho}.
\end{eqnarray}
In the present case,  we deal with symmetric states of $N_c$ quarks which means that we have to take $[f] = [N_c]$. 
The Casimir operator is $C^{SU(6)}_{[N_c]} = \frac{5}{12}N_c(N_c + 6)$. The sum over $\rho$, $\lambda''$
and $\mu''$ contains isoscalar factors derived in Ref. \cite{Matagne:2006xx} and presented in Table I of that reference.

For symmetric states $[f] = [N_c]$ it happens that the sum over $S''$ is such that $F(N_c,S,S')$ becomes $\mathcal{O}(N^0_c)$,
and the order of $0_6$ is given by the factor $\frac{1}{N_c}$. In other words it means that the three-flavor case is more
subtle than the two-flavor case, as already pointed out in Ref. \cite{JENKINS}, because $G^{ia}$ does not have the same 
$N_c$ dependence in the flavor weight diagram.

\section{Mixing of quartets and doublets}\label{Mixing}

Mixing between doublets and quartets with the same $J$ is  possible due to off-diagonal 
matrix elements of $O_2$ and $O_6$. Introducing a mixing angle $\theta_J$ one can define

\begin{eqnarray}
|N_J(\mathrm{upper}) \rangle = \cos \theta_J |^4N_J \rangle +
 \sin \theta_J |^2N_J \rangle, \nonumber \\
|N_J(\mathrm{lower}) \rangle =  
- \sin \theta_J |^4N_J \rangle +  \cos \theta_J |^2N_J \rangle ,
\end{eqnarray}
where  $|^4N_J \rangle$ and $|^2N_J \rangle$ are the theoretical states 
used in the fit and ${\it upper}$ and ${\it lower}$ are   the physical states 
with upper and lower energies. One can give an estimate of the mixing angle $\theta_J$
to order $\mathcal{O}(N^0_c)$ which can be obtained from a simplified mass formula
including $O_1$, $O_2$ and $O_6$, the first being order $\mathcal{O}(N_c)$, the two others order $\mathcal{O}(N^0_c)$.
In Table \ref{Off-diag}
we exhibit the off-diagonal matrix elements of $O_2$ and $O_6$ as a function of $N_c$. 
Like the diagonal matrix elements, they are also order 
$\mathcal{O}(N^0_c)$. 

\begin{table*}[h!]
\begin{center}
\caption{Off-diagonal matrix elements of  $O_2$ and $O_6$ for all states belonging to the 
$[{\bf 70},2^+]$ multiplet.  }
\label{Off-diag}
\renewcommand{\arraystretch}{2.3} {\scriptsize
\begin{tabular}{lcccccccccc}
\hline
\hline
   &\hspace{0cm} $O_2$ 
  & \hspace{0cm}  $O_6$  
  &\hspace{0cm}     \\
  \hline  
$^28_{\frac{3}{2}} -$ $^48_{\frac{3}{2}}$  & $-\sqrt{\frac{N_c+3}{4N_c}}$ &
$-\frac{7}{8}\sqrt{\frac{N_c+3}{N_c}}$ & \\
$^28_{\frac{5}{2}}-$ $^48_{\frac{5}{2}}$  & $-\frac{1}{3}\sqrt{\frac{7(N_c+3)}{2N_c}}$ & 
$\sqrt{\frac{7}{32}}\sqrt{\frac{N_c+3}{N_c}}$ & \\
\hline \hline
\end{tabular}}
\end{center}
\end{table*}
Then we have to take $N_c \to \infty$ in all matrix elements of $O_2$ and $O_6$ \cite{COLEB1}.
As an example we show here 
the matrix of the $N_{5/2}$ states
\begin{eqnarray}
\label{fivehalfs}
M^{\ell=2}_{N_{5/2}} & = &
 \left( 
\renewcommand{\arraystretch}{1.5}
 \begin{array}{cc}
c_1 N_c + \frac{2}{3} c_2  &~~~~~~  \sqrt{\frac{7}{2}}~(- \frac{1}{3}c_2 + \frac{1}{4} ~c_6 )\\
\sqrt{\frac{7}{2}}~(- \frac{1}{3}c_2 + \frac{1}{4} ~c_6 ) &~~~~~~   c_1 N_c - \frac{1}{6} ~c_2 + \frac{5}{8} ~c_6 
\end{array} 
\right), 
\end{eqnarray}   
This suggests that the general form of a 2 $\times$ 2 matrix to be diagonalized is
\begin{eqnarray}
\label{N}
M^{\ell}_{N_{J}} & = &
 \left( 
 \begin{array}{cc}
A &~~~~~~ B \\
B  &~~~~~~ C
\end{array} 
\right), 
\end{eqnarray}  
so the mixing angle turns out to be
\begin{equation}      
\tan 2 \theta = - \frac{2B}{C-A}.           
\end{equation}    
Replacing $A, B$ and $C$ by their values from Eq. (\ref{fivehalfs}), one obtains
\begin{equation}
 \tan 2\theta_{5/2} = -\frac{\sqrt{\frac{7}{2}} (-\frac{2}{3}c_2 + \frac{1}{2}c_6)}{-\frac{5}{6}c_2 + \frac{5}{8}c_6 }
\end{equation}
Using the coefficients  $c_2$ and $c_6$ from Table \ref{operators} one obtains $\tan 2\theta_{5/2} \approx$ -1.49,
which gives  $\theta_{5/2}\approx$ - 28 degrees.
This is quite a large mixing angle. In the real
case, one has to introduce corrections of order $1/N_c$. However the mixing angles are completely unknown 
experimentally for the $[{\bf 70},2+]$, contrary to the  $[{\bf 70},1^-]$ multiplet (for the most recent analysis
see Ref. \cite{Willemyns:2015hgy}). 
So, a comparison between theory and experiment is not yet possible.

\section{Breaking operators}\label{BREAK}

Here we present some details of the calculation of the diagonal matrix elements of the SU(3) breaking operators 
$B_2$ and $B_3$ presented in Table \ref{break70}. In the context of our approach, where the baryon is treated as a system of $N_c$ 
quarks irrespective of its spin-flavor symmetry, they are defined as 
\begin{equation}
B_2 = \frac{1}{N_c}(L^i G^{i8} - \frac{1}{2 \sqrt{3}} L \cdot S),
\end{equation}
and 
\begin{equation}
B_3 =\frac{1}{N_c}(S^iG^{i8} - \frac{1}{2 \sqrt{3}} S\cdot S),
\end{equation}
where the angular momentum operator $L^i$ acts on the entire system of  $N_c$ quarks.
The matrix elements of the operators $L^i G^{i8}$ and $S^iG^{i8}$ have been obtained in Ref. \cite{Matagne:2011fr}.
Their analytic expressions are
\begin{eqnarray}\label{LG}
 \lefteqn{\langle 
 (\lambda'\mu')Y'I'I_3';\ell'S' JJ_3|(-1)^{i}L^iG^{-i8}|(\lambda\mu)YII_3;\ell S JJ_3\rangle 
 =}\nonumber \\ & &  \delta_{\ell'\ell} 
 (-1)^{\ell+S'+J}\sqrt{\frac{C^{SU(6)}_{[f]}}{2}}\sqrt{\ell(\ell+1)
(2\ell+1)
(2S'+1)}
 \left\{\begin{array}{ccc}
   S' & \ell & J \\
   \ell & S & 1
  \end{array}\right\} \nonumber \\
  & & \times \sum_{\rho}\left(\begin{array}{cc||c}
	 (\lambda\mu) & (11)  & (\lambda'\mu')\\
	 YI           &  00   &  YI
        \end{array}\right)_{\rho}
  \left(\begin{array}{cc||c}
         [f]           & [21^4] & [f] \\
	 (\lambda\mu)S & (11)1  & (\lambda'\mu')S'
        \end{array}\right)_{\rho},
\end{eqnarray}
and
\begin{eqnarray}\label{SG}
 \lefteqn{\langle 
 (\lambda'\mu')Y'I'I_3';\ell'S' JJ_3|(-1)^{i}S^iG^{-i8}|(\lambda\mu)YII_3;\ell S JJ_3\rangle 
 =   \delta_{\ell'\ell} \delta_{S'S}}\nonumber \\
  & &\times \sqrt{\frac{C^{SU(6)}_{[f]}}{2}}\sqrt{S(S+1)}
 \sum_{\rho}\left(\begin{array}{cc||c}
	 (\lambda\mu) & (11)  & (\lambda'\mu')\\
	 YI           &  00   &  YI
        \end{array}\right)_{\rho}
  \left(\begin{array}{cc||c}
         [f]           & [21^4] & [f] \\
	 (\lambda\mu)S & (11)1  & (\lambda'\mu')S
        \end{array}\right)_{\rho},
\end{eqnarray}
respectively. Note that the factor $\sqrt{\ell(\ell+1)}$ appearing in Eq. (\ref{LG}) is missing in Eq. (D4)
of Ref. \cite{Matagne:2011fr}.

The matrix elements of $S^i G^{i8}$ for mixed symmetric  $[{{\bf 70}, 2^+}]$ states
have been straightforwardly obtained  from the analytic forms of the matrix 
elements of $S^i G^{i8}$ exhibited in Table X of Ref. \cite{Matagne:2011fr},
where they were derived in the context of the
multiplet $[{{\bf 70}, 1^-}]$, but they can be applied to any angular momentum $\ell$ and  parity.
These analytic forms are simple ratios of polynomials in the variables $N_c$, the isospin $I$  
and the strangeness $\mathcal{S}$.

From the definitions (\ref{LG}) and (\ref{SG}) one can see that the expectation values of
$L^i G^{i8}$ and $S^i G^{i8}$ are related by
\begin{equation}
\langle L^i G^{i8} \rangle =
\delta_{S'S} 
 (-1)^{\ell+S'+J}\sqrt{\frac{\ell(\ell+1)(2\ell+1)(2S'+1)}{S(S+1)}}
 \left\{\begin{array}{ccc}
   S' & \ell & J \\
   \ell & S & 1
  \end{array}\right\} 
\langle S^i G^{i8} \rangle, 
\end{equation}
which can help to easily find the entries for $L^i G^{i8}$ using $S^i G^{i8}$ from Table X of Ref. \cite{Matagne:2011fr}
and the corresponding 6j coefficients.
Note also that
the expectation values of $S^i G^{i8}$ are independent of $J$, as seen from Table X of Ref. \cite{Matagne:2011fr}.
The expectation values of $L^i G^{i8}$ for $J'$ and $J$ at fixed $S$    can be obtained from the ratio of
the corresponding $6j$  coefficients.
\begin{equation}\label{otherJ}
\frac{\langle L^i G^{i8} \rangle_{J'}}{\langle L^i G^{i8} \rangle_{J}} = (-1)^{J'-J}
\frac{\left\{\begin{array}{ccc}
   S & \ell & J' \\
   \ell & S & 1
  \end{array}\right\}}
{\left\{\begin{array}{ccc}
   S & \ell & J \\
   \ell & S & 1
  \end{array}\right\}}.
\end{equation}
The  use of  (\ref{otherJ}) can simplify the calculation of $B_2$.


\begin{thebibliography}{9}


\bibitem{'tHooft:1973jz}
  G.~'t Hooft,
  Nucl.\ Phys.\ B {\bf 72} (1974) 461.


\bibitem{Witten:1979kh}
  E.~Witten,
  Nucl.\ Phys.\ B {\bf 160} (1979) 57.

\bibitem{Gervais:1983wq}
  J.~L.~Gervais and B.~Sakita,
  Phys.\ Rev.\ Lett.\  {\bf 52} (1984) 87;
  Phys.\ Rev.\ D {\bf 30} (1984) 1795.

\bibitem{DM93}
R.~Dashen and A.~V.~Manohar,
Phys.~Lett. {\bf B315}, 425 (1993);
Phys.~Lett. {\bf B315}, 438 (1993).

\bibitem{Cook:1965qu}
  T.~Cook, C.~J.~Goebel and B.~Sakita,
  Phys.\ Rev.\ Lett.\  {\bf 15} (1965) 35.


\bibitem{Bardakci:1983ev}
  K.~Bardakci,
  Nucl.\ Phys.\ B {\bf 243} (1984) 197.


\bibitem{Jenk1}
E.~Jenkins, Phys.~Lett.  {\bf B315}, 441 (1993).

\bibitem{DJM94}
R.~Dashen, E.~Jenkins, and A.~V.~Manohar,
Phys.~Rev.  {\bf D49}, 4713 (1994).


\bibitem{DJM95}
R. Dashen, E. Jenkins, and A. V. Manohar,
Phys. Rev. {\bf D51}, 3697 (1995).

\bibitem{CGO94}
C.~D.~Carone, H.~Georgi and S.~Osofsky,
Phys.~Lett. {\bf B322}, 227 (1994);\\
M.~A.~Luty and J.~March-Russell,
Nucl.~Phys. {\bf B426}, 71 (1994);\\
M.~A.~Luty, J.~March-Russell and  M.~White,
Phys.~Rev.  {\bf D51}, 2332 (1995).

\bibitem{JL95}
E.~Jenkins and R.~F.~Lebed,
Phys.~Rev.  {\bf D52}, 282 (1995).

\bibitem{DDJM96}
J.~Dai, R.~Dashen, E.~Jenkins, and A.~V.~Manohar,
Phys.~Rev.   {\bf D53}, 273 (1996).
\bibitem{Goi97}
J.~L. Goity, Phys. Lett. B {\bf 414}, 140 (1997). 

\bibitem{Edwards:2012fx}
  R.~G.~Edwards, N.~Mathur, D.~G.~Richards and S.~J.~Wallace,
  Phys.\ Rev.\ D {\bf 87} (2013) 054506.


\bibitem{PY} 
D. Pirjol and T.~M. Yan,
 Phys. Rev.  {\bf D57}, 1449 (1998); ibid. {\bf D57}, 5434 (1998).


 \bibitem{COLEB1}
  T.~D.~Cohen and R.~F.~Lebed,
  Phys.\ Rev.\ Lett.\  {\bf 91}, 012001 (2003); 
  Phys. Rev. {\bf D67} (2003) 096008.


\bibitem{GSS03}
J.~L. Goity,   C.~L.~Schat and N.~N.~Scoccola,
 Phys. Lett. B {\bf 564}, 83 (2003).




\bibitem{Matagne:2004pm}
  N.~Matagne and F.~Stancu,
  Phys.\ Rev.\ D {\bf 71} (2005) 014010.

\bibitem{CCGL}
C.~E. Carlson, C.~D. Carone, J.~L. Goity and R.~F. Lebed,
 Phys. Lett. {\bf B438}, 327  (1998);
 Phys. Rev. {\bf D59}, 114008 (1999).




\bibitem{Matagne:2006zf}
  N.~Matagne and F.~Stancu,
  Phys.\ Rev.\ D {\bf 74} (2006) 034014.




\bibitem{Matagne:2006dj}
  N.~Matagne and F.~Stancu,
  Nucl.\ Phys.\ A {\bf 811} (2008) 291.

\bibitem{Matagne:2008kb}
  N.~Matagne and F.~Stancu,
  Nucl.\ Phys.\ A {\bf 826} (2009) 161.

\bibitem{Matagne:2014lla}
  N.~Matagne and F.~Stancu,
  Rev.\ Mod.\ Phys.\  {\bf 87} (2015) 211.

\bibitem{PDG} 
K.~A.~Olive et al. (Particle Data Group)  Chin.\ Phys.\  {\bf  C38} (2014) 090001.



\bibitem{Crede:2013kia}
  V.~Crede and W.~Roberts,
  Rept.\ Prog.\ Phys.\  {\bf 76} (2013) 076301.



\bibitem{Carlson:2000zr}
  C.~E.~Carlson and C.~D.~Carone,
  Phys.\ Lett.\ B {\bf 484} (2000) 260.
  
\bibitem{Matagne:2005gd}
  N.~Matagne and F.~Stancu,
  Phys.\ Lett.\  B {\bf 631} (2005) 7.






\bibitem{Matagne:2013cca}
  N.~Matagne and Fl.~Stancu,
  Phys.\ Rev.\ D {\bf 87} (2013) 076012.





\bibitem{Semay:2007cv}
  C.~Semay, F.~Buisseret, N.~Matagne and F.~Stancu,
  Phys.\ Rev.\  D {\bf 75} (2007) 096001.


\bibitem{Semay:2007ff}
  C.~Semay, F.~Buisseret and Fl.~Stancu,
  Phys.\ Rev.\ D {\bf 76} (2007) 116005.


\bibitem{Buisseret:2008tq}
  F.~Buisseret, C.~Semay, Fl.~Stancu and N.~Matagne,
  (Bled Workshops in Physics, Vol.9 No.1, p.9)
  arXiv:0810.2905 [hep-ph].




%


\bibitem{PDG2002}
K. Hagiwara et al. (Particle Data Group)  Phys.\ Rev.\  {\bf  D66} (2002) 010001.



\bibitem{Anisovich:2011fc}
  A.~V.~Anisovich, R.~Beck, E.~Klempt, V.~A.~Nikonov, A.~V.~Sarantsev and U.~Thoma,
  Eur.\ Phys.\ J.\ A {\bf 48} (2012) 15.


\bibitem{Matagne:2011fr}
  N.~Matagne and Fl.~Stancu,
  Phys.\ Rev.\ D {\bf 83} (2011) 056007.


\bibitem{Matagne:2012vq}
  N.~Matagne and F.~Stancu,
  Phys.\ Rev.\ D {\bf 86} (2012) 076007.





\bibitem{Cohen:2003fv}
  T.~D.~Cohen, D.~C.~Dakin, A.~Nellore and R.~F.~Lebed,
  Phys.\ Rev.\ D {\bf 69} (2004) 056001.


\bibitem{Matagne:2006xx}
  N.~Matagne and Fl.~Stancu,
  Phys.\ Rev.\ D {\bf 73} (2006) 114025.







\bibitem{Cohen:2005ct}
  T.~D.~Cohen and R.~F.~Lebed,
  Phys.\ Rev.\ D {\bf 72} (2005) 056001.




\bibitem{Stassart:1997vk}
  P.~Stassart and F.~Stancu,
  Z.\ Phys.\ A {\bf 359} (1997) 321.

\bibitem{Ablikim:2009iw}
  M.~Ablikim {\it et al.} [BES Collaboration],
  Phys.\ Rev.\ D {\bf 80} (2009) 052004.




\bibitem{Pirjol:2003ye}
D.~Pirjol and C.~Schat, Phys. Rev. {\bf  D67}, 096009 (2003).

\bibitem{Arndt:2006bf}
  R.~A.~Arndt, W.~J.~Briscoe, I.~I.~Strakovsky and R.~L.~Workman,
  Phys.\ Rev.\ C {\bf 74} (2006) 045205.


\bibitem{Litchfield:1971ri}
  P.~J.~Litchfield {\it et al.},
  Nucl.\ Phys.\ B {\bf 30} (1971) 125.



\bibitem{Zhang:2013sva}
  H.~Zhang, J.~Tulpan, M.~Shrestha and D.~M.~Manley,
  Phys.\ Rev.\ C {\bf 88} (2013) 3,  035205.




\bibitem{Gopal:1980ur}
  G.~P.~Gopal,
  RL-80-045, C80-07-14.1-9.


\bibitem{Biagi:1986vs}
  S.~F.~Biagi {\it et al.},
  Z.\ Phys.\ C {\bf 34} (1987) 175.











\bibitem{Gopal:1976gs}
  G.~P.~Gopal {\it et al.} [Rutherford-London Collaboration],
  Nucl.\ Phys.\ B {\bf 119} (1977) 362.




\bibitem{Goity:2007sc}
  J.~L.~Goity and N.~Matagne,
  Phys.\ Lett.\  B {\bf 655}, 223 (2007).


\bibitem{Matagne:2012tm}
  N.~Matagne and F.~Stancu,
  Phys.\ Rev.\ D {\bf 85} (2012) 116003.


\bibitem{JENKINS} E. Jenkins, Ann. Rev. Nucl. Sci. {\bf 48}, 81 (1998).



\bibitem{Willemyns:2015hgy}
  C.~Willemyns and C.~Schat,
  Phys.\ Rev.\ D {\bf 93} (2016) 3,  034007.



\end{thebibliography}
\end{document}